\documentclass[3p,times,twocolumn]{elsarticle}

\usepackage{ecrc}

\volume{00}

\firstpage{1}

\journalname{Nuclear and Particle Physics Proceedings}

\runauth{M. Kordell, A. Majumder}

\jid{nppp}

\jnltitlelogo{Nuclear and Particle Physics Proceedings}

\usepackage{amssymb}
\usepackage{amsmath}

\usepackage[figuresright]{rotating}

\begin{document}

\begin{frontmatter}



\dochead{}

\title{Event-by-Event Simulations of Jet Modification Using the MATTER Event Generator}


\author{Michael C. Kordell II and Abhijit Majumder}
\address{Wayne State University, Detroit, MI 48202, USA}

\begin{abstract}
The modification of hard jets in the Quark Gluon Plasma (QGP) is studied using the MATTER event generator. Based on the higher twist formalism of energy loss, the MATTER event generator simulates the evolution of highly virtual partons through a medium. These partons sampled from an underlying PYTHIA kernel undergo splitting through a combination of vacuum and medium induced emission. The momentum exchange with the medium is simulated via the jet transport coefficient $\hat{q}$, which is assumed to scale with the entropy density at a given location in the medium. The entropy density is obtained from a relativistic viscous fluid dynamics simulation (VISH2+1D) in 2+1 space time dimensions. Results for jet and hadron observables are presented using an independent fragmentation model. These proceedings will focus on the physics input and simulation details of the MATTER event generator as compared to a variety of test observables.
\end{abstract}

\begin{keyword}
Heavy-ion \sep Jet \sep Nuclear modification \sep Monte-Carlo

\end{keyword}

\end{frontmatter}


\section{Introduction}

We study heavy-ion collisions to examine the properties of the Quark Gluon Plasma (QGP).  The QGP is a form of matter characterized by deconfined quarks and gluons that we expect to appear in these collisions.  One of the methods available to study the QGP is by determining the modification of jets due to this medium.  A jet, for the purpose of this discussion, is a shower of particles originating from a high transverse momentum ($p_T$) parton.

In order to compare analytical approaches of jet modification to experimental data, a Monte-Carlo event generator is extremely useful due to the ability to directly simulate the physical process and to make "measurements" from the simulation, formulated in very close analogy to experimental observables.  There are several other preexisting simulations; Q-PYTHIA \cite{Armesto:2009fj} which is based on the Armesto-Salgado-Wiedemann (ASW) scheme \cite{Armesto:2003jh} and MARTINI \cite{Schenke:2009gb} which is based on the Arnold-Moore-Yaffe (AMY) scheme \cite{Qin:2007rn}.  There are also a number of event generators that are not strictly based on analytical models, including JEWEL \cite{Zapp:2008gi}\cite{Zapp:2012ak}, YaJEM \cite{Renk:2009hv}\cite{Renk:2012ve}, and PYQUEN \cite{Lokhtin:2005px} which include medium effects by manually modifying various matrix elements.  Generally, these simulations have handled the inclusion of a medium by taking a vacuum event generator and to either add the modification of the jet due to the medium on top of a full vacuum shower, or to alter the vacuum shower generation in such a way that both vacuum radiation and medium induced radiation are performed concurrently.

However, in addition to the technical construction of including a medium, there are two issues that should be considered when adding medium effects to a simulation.  The first of which is constructing the space-time structure of the shower since the medium itself has a space-time structure.  The second is a modification of hadronization; the shower partons can potentially recombine with partons from the thermal medium.  These two issues are not addressed in a copacetic fashion in the schemes mentioned previously, though we note that YaJEM has phenomenologically incorporated fluctuations in space-time structure \cite{Renk:2009nz}.

The event generator we present in this discussion is based on the Higher-Twist scheme \cite{Majumder:2009ge}\cite{Majumder:2008zg}\cite{Majumder:2011uk}\cite{Majumder:2010qh}.  We also attempt to include a consistent space-time structure with fluctuations within the simulation.  The Higher-Twist scheme itself is applicable to high energy, high virtuality partons, in contrast to the other schemes (AMY and ASW) that are more applicable to lower virtuality (though still high energy) partons.  Our simulation is constructed with PYTHIA \cite{Sjostrand:2014zea} to sample the initial high $p_T$ parton, the OSU (Ohio State University) hydrodynamic simulation \cite{Shen:2014vra} to provide the thermal medium, and the MATTER event generator \cite{Majumder:2013re} for jet quenching.  In the remainder of these proceedings, we will briefly discuss the Higher-Twist model, some details of this event generator, and present some preliminary results from the simulation compared to experimental data.

\section{Simulation}

The simulation begins in two parts: PYTHIA and the OSU hydrodynamic simulation.  PYTHIA was used to generate the initial hard parton for the shower.  Nuclear shadowing was not included, but is a planned future modification. PYTHIA was setup with the center-of-mass energy of the collision and the $p_T$ bounds for the hard process as well as turning off final state radiation and all hadron level processes.  From the produced event, the leading two partons at midrapidity ($y \leq \pm 0.25$) were taken for jet quenching.  Multiple hard-$p_{T}$ bins were used to enhance statistics for high $p_T$ values ($\geq 5 GeV$) to compenstate for the falling jet spectrum.  The cross-section for each of these bins was obtained from PYTHIA for use in subsequent calculations.

The OSU hydrodynamic code is an event-by-event 2+1d hydrodynamic simulation for relativistic heavy-ion collisions with fluctuating initial conditions.  It was used to generate both a medium for jet quenching and to report the initial density profile $T_{AA}(x,y)$ for sampling the hard parton's initial location.  While it could have been used to generate thermal partons and/or hadrons for the background of the jet, this was not performed for this analysis.

\begin{figure}[htb]
	\includegraphics[width=0.45\textwidth]{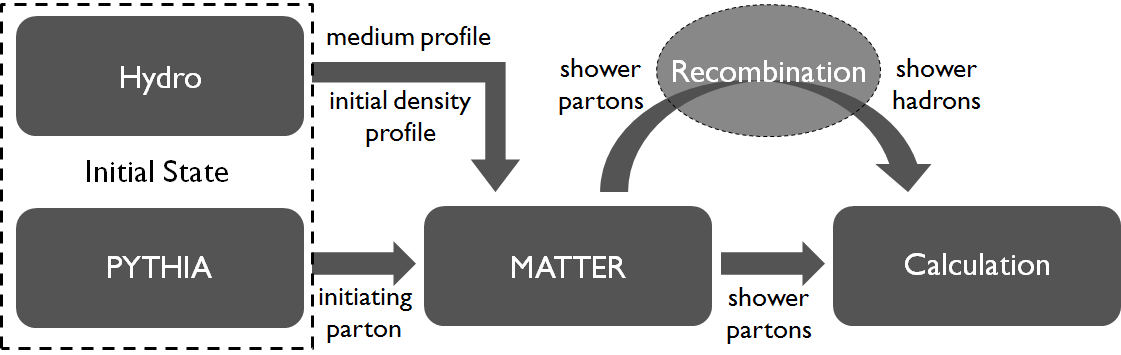}
	\caption{Code flowchart showing data input/output for each code set}
	\label{fig1}
\end{figure}

The hard parton from PYTHIA and the medium from the hydrodynamic simulation were then read by the MATTER event generator for jet showering.  The MATTER event generator was used to simulate jet showering in both vacuum and in medium.  While the parton showers generated by MATTER could be be used with a hadronization scheme, such as Texas A\&M's recombination code\cite{Fries:2016gjx} for hadronic results, we instead generated partonic spectra and used the Kniehl-Kramer-Potter (KKP) fragmentation function\cite{Kniehl:2000fe} to generate leading hadron spectra.  In addition, Fastjet was run over the generated partonic showers to generate full jets using the anti-$k_T$ algorithm with an $R=0.4$  for analysis.  A general flowchart of this simulation is shown in Fig.~\ref{fig1}.

\section{The MATTER Event Generator}
The jet quenching portion of the simulation was performed with the MATTER++ (Modular All Twist Transverse scattering based Energy-loss Routines in C++) event generator \cite{Majumder:2013re}.  It is based on the Higher-Twist formalism \cite{Majumder:2013re} and as such it is primarily applicable to the high energy, high virtuality portion of a particular jet in the 'few' scatterings (meaning zero to one) per emission limit. In this regime, light quark modification is sensitive to the high $Q^{2}$, low-x part of the in-medium gluon distribution.  In order to introduce space-time into the shower, the notion that the uncertainty in the momentum is conjugate to the position (and likewise, that the uncertainty in the position is conjugate to the momentum) was used.  For a reasonable uncertainty, we assert that $\delta q^+ << q^+$. We assume a Gaussian distribution around $q^{+}$ and insist that 

\begin{equation} <\tau> = 2q^{-} / Q^{2} .\end{equation}

Then to obtain the $z^-$ we assume a $\delta q^{+}$ distribution.  Thus, we obtain:

\begin{equation} \rho(\delta q^{+}) = \frac{exp \left [-\frac{(\delta q^{+})^{2}}{2[2(q^{+}) / \pi]} \right ] }{\sqrt{2 \pi [2(q^{+})^{2} / \pi]}} \ \ .\end{equation}

The off-shell quark will have mometum $q=[q^-,q^+,0,0]$.  This allows for the parton's travel length to the next split to be determined.

Before the length traversed for the current parton can be calculated however, its virtuality must first be determined.  This is done by sampling the Sudakov form factor to obtain the maximum virtuality $\mu^2$ (which is also the running scale) of the splitting parton, which is constructed as:

\begin{multline} S_{\xi}(Q^2_0,Q^2) = exp \bigg[ \int^{Q^2}_{2Q^2_0} \frac{d\mu^2}{\mu^2} \frac{\alpha_{s}( \mu^2 )}{2\pi} \cdot \\
\int^{1-Q_{0}/Q}_{Q_{0}/Q} dy P_{qg}(y) \bigg  \{ 1+ \int^{\xi^-_i + \tau^-}_{\xi^-_i} d\xi K_{p^- , \mu^2} \bigg  \} \bigg ] \ \ . \end{multline}

The Sudakov itself gives the probability of the parton having no emission from initial virtuality $2Q_0$ to final virtuality $Q$.  $P_{qg}(y)$ is the splitting function for a quark to split into a quark and a gluon where the final quark carries momentum $yq^-$ and the gluon carries momentum $(1-y)q^-$.  The single emission, multiple scattering kernel $K$ as a function of the momentum fraction $y$ and the location of the parton $\xi$ starting from location $\xi_i$ is:

\begin{equation} K_{p^- , \mu^2}(y, \xi) = \frac{2 \hat{q} }{\mu^2} \bigg [2 - 2cos \bigg \{ \frac{\mu^2 (\xi - \xi_i)}{2p^- y(1-y)} \bigg \} \bigg ]  \ \ ,\end{equation}

\noindent
where $\hat{q}$ is the jet transport coefficient \cite{Baier:2002tc}\cite{Majumder:2012sh}.

Since this simulation is based on the Higher-Twist scheme, multiple emissions are ordered in $p_T$.  These ordered multiple emissions are only considrered when the multiple soft scatterings mildly effect the virtuality of the parent parton.  For partons where the virtuality has become too low, this calculation is no longer applicable.  This means that our procedure is only valid while:

\begin{equation} \frac{\hat{q}\tau}{\mu^{2}} \lesssim 1 \ \ .\end{equation}

With this, the code can read in a high-$p_T$ parton, that was generated using PYTHIA, and begin to generate the shower.  In order to do so the entropy density of the medium is read in from the pre-run hydrodynamic simulation.  This is used to modify $\hat{q}$ in the Sudakov form factor.  The Sudakov is sampled to return the largest virtuality allowable for the process.  This virtuality is then used to determine the distance travelled by the parton before it splits.  The splitting function is sampled to determine the mometum fraction $y$ of one of the outgoing partons.  This process is repeated over the outgoing partons for each iteration until all the generated partons have a virtuality at or less than 1 GeV$^2$, beyond which the Higher Twist formalism is no longer applicable.

The final partons in the shower are then checked to determine if they are able to escape the medium; this is done by removing any parton that is further than 1fm from the edge of the medium.  This can remove high energy but low virtuality partons, though this is a rare occurance; instead a planned method to deal with these partons is by handing them off to an event generator that includes a multiple scatterings per emission treatment, such as MARTINI \cite{Schenke:2009gb}.  The low energy, low virtuality partons that were removed are planned to be used to generate source terms for a medium response to the jet.  The partons that escape the medium are then taken as the final generated shower from the MATTER code.

\section{Preliminary Results}

Jets had been produced over a range of hard $p_{T}$ bins: 2.5-52.5GeV in 5GeV wide bins for Au+Au and 2.5-227.5GeV for Pb+Pb.  The calculation for the nuclear modification factor $R_{AA}$ started by taking each of these hard pT bins, weighting it by its corresponding p+p cross-section, then summing over all the aforementioned hard $p_{T}$ bins to get the total spectra.  The nuclear modification factor $R_{AA}$ is defined as

\begin{eqnarray}
R_{AA} = \frac{ \int\limits_{b_{min}}^{b_{max}} d^2 b \frac{d^4N_{AA}}{d^2 p_T d y d^2 b} }
{  \langle N_{bin} (b_{min}, b_{max})\rangle    \frac{d^3 N_{pp}}{ d^2 p_T dy}  } \ \ ,
\end{eqnarray}

\noindent
where N is the yield of jets or leading hadrons binned in $p_T$ and rapidity.  The numerator in the preceeding formula also includes an integral over a range of the impact parameter $b$ to construct bins in centrality.  The factor $\langle N_{bin} (b_{min}, b_{max})\rangle$ is the mean number of binary nucleon-nucleon collisions $N_{bin}$ of a nuclear collision from a given centrality bin for $b_{min} < b < b_{max}$.  This gives a method of quantifying nuclear effects, as it allows for a comparison of heavy-ion data where the QGP was created to p+p events where we do not expect the presence of the QGP or other nuclear influcences such as initial state cold nuclear matter effects.

The A+A cross-section in this case is just $N_{bin}$ multiplied by the previous p+p cross-section. The KKP fragmentation function was applied to partonic spectra to obtain leading hadron (or pion) spectra, which were then used to calculate $R_{AA}$ as mentioned above.  The results of these calculations are given in Figures 2 and 3.  These results use a $\hat{q}_0 = 2.4 GeV^2/fm$ except for leading hadron data from A+A collisions; $\hat{q}_0$ is the value of $\hat{q}$ at the center of an averaged 0-5\% centrality bin Au+Au collision.

\begin{figure}[htb]
	\includegraphics[width=0.45\textwidth]{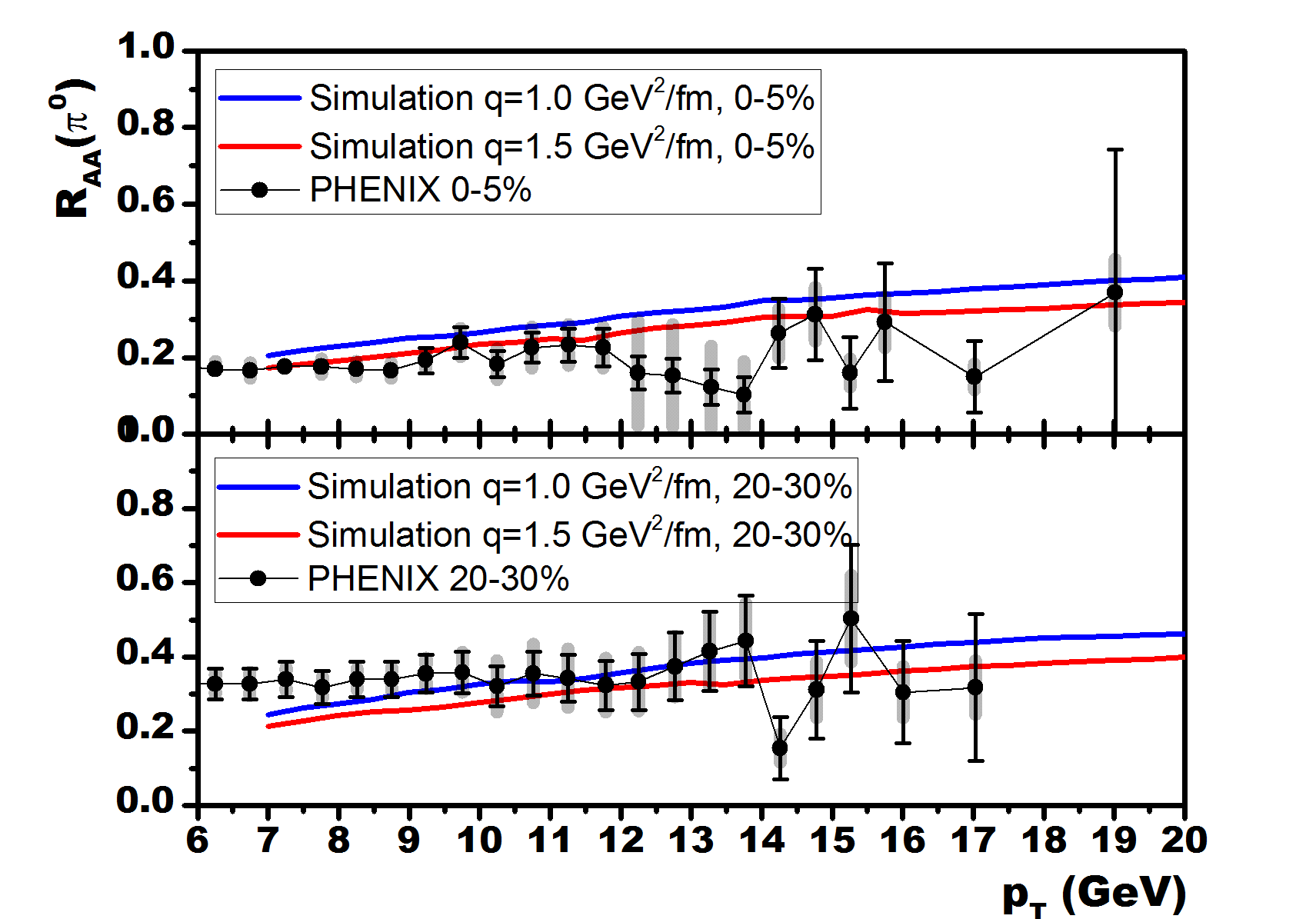}
	\caption{Leading pion $R_{AA}$ for 200 GeV Au+Au compared to PHENIX data \cite{Adare:2012wg} for varying $\hat{q}_0$}
	\label{fig2}
\end{figure}%

\begin{figure}[htb]
	\includegraphics[width=0.45\textwidth]{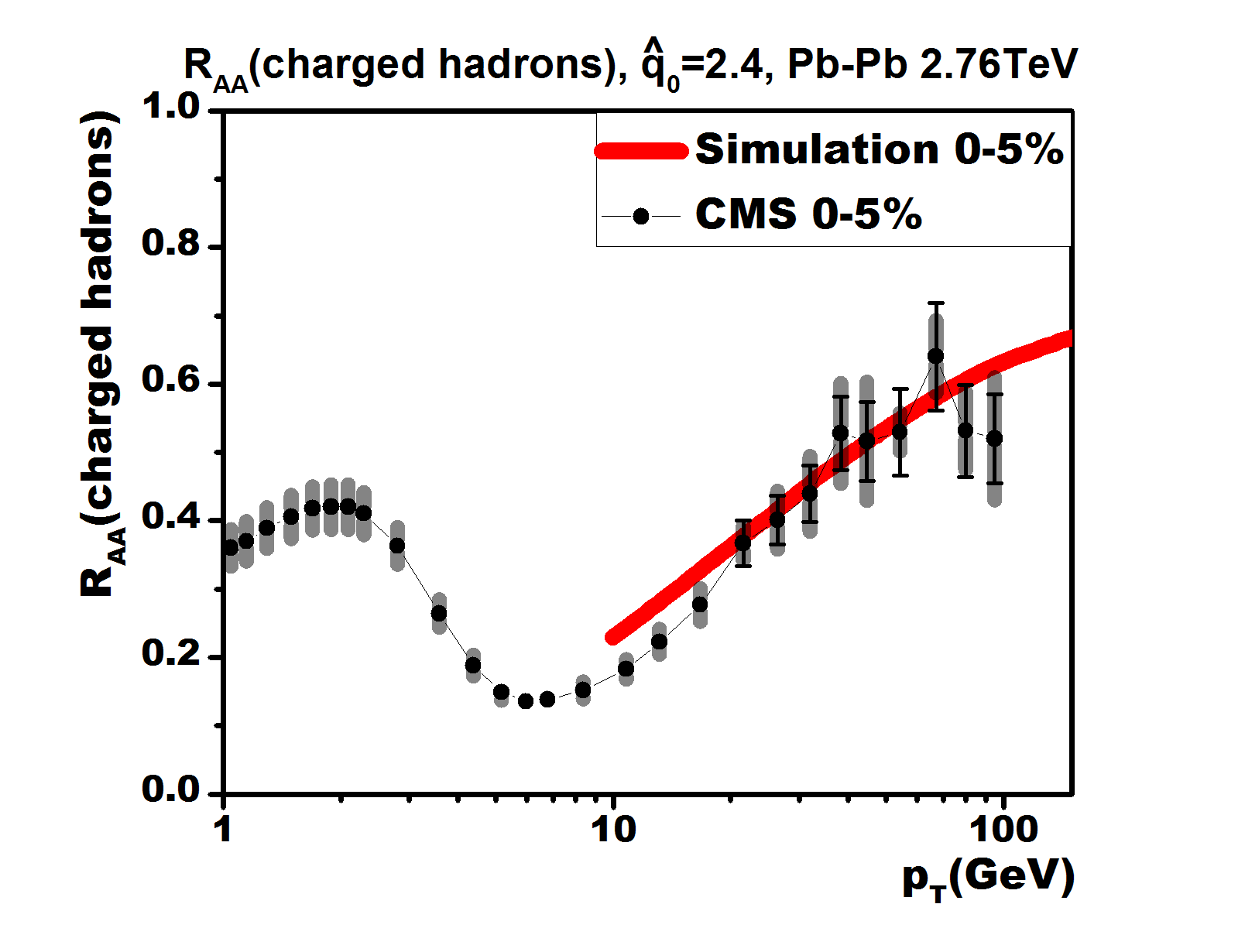}
	\caption{Leading hadron $R_{AA}$ for 2.76 TeV Pb+Pb compared to CMS data \cite{CMS:2012aa} with $\hat{q}_0 = 2.4 GeV^2/fm$}
	\label{fig3}
\end{figure}%

\section{Conclusion}

As the plots above show, the simulation produces results that are consistent with experimental data.  In the near future, we intend to further refine the presented results.  We also plan to present further analyses including $v_{2}$ and jet shapes.  Further plans include a method of handling partons with a virtuality of 1 GeV or less, incorporating medium response via a source term, including thermal hadrons, incorporating the Texas A\&M recombination code, and including thermal-shower recombination hadrons.  With the execution of these plans we hope to see even better agreement with experimental data, to compare to more experimental data, and to predict a number of future experimental results.

This work was supported in part by the NSF under grant number PHY-1207918 and by the U.S. DOE under grant number DE-SC0013460.







\end{document}